\documentclass[acmtog, nonacm]{acmart}

\usepackage{booktabs} 

\usepackage[ruled]{algorithm2e} 

\SetAlFnt{\small}
\SetAlCapFnt{\small}
\SetAlCapNameFnt{\small}
\SetAlCapHSkip{0pt}

\acmJournal{TOG}

\begin{document}
\title{Geodiffussr: Generative Terrain Texturing with Elevation Fidelity}

\author{Tai Inui}
\orcid{0009-0000-2387-2790}
\affiliation{%
 \institution{Waseda University}
 \city{Tokyo}
 \country{Japan}}
\affiliation{%
 \institution{Rikka Inc.}
 \city{Tokyo}
 \country{Japan}}

\author{Alexander Matsumura}
\affiliation{%
 \institution{Waseda University}
 \city{Tokyo}
 \country{Japan}}

\author{Edgar Simo-Serra}
\orcid{0000-0003-2544-8592}
\affiliation{%
 \institution{Waseda University}
 \city{Tokyo}
 \country{Japan}}

\renewcommand\shortauthors{Inui et al}

\begin{abstract}
Large-scale terrain generation remains a labor-intensive task in computer graphics. We introduce \textit{Geodiffussr}, a flow-matching pipeline that synthesizes text-guided texture maps while strictly adhering to a supplied Digital Elevation Map (DEM). The core mechanism is multi-scale content aggregation (MCA): DEM features from a pretrained encoder are injected into UNet blocks at multiple resolutions to enforce global-to-local elevation consistency. Compared with a non-MCA baseline, MCA markedly improves visual fidelity and strengthens height–appearance coupling (FID $\downarrow$ 49.16\%, LPIPS $\downarrow$ 32.33\%, $\Delta$dCor $\downarrow$ to 0.0016). To train and evaluate Geodiffussr, we assemble a \emph{globally distributed, biome- and climate-stratified} corpus of triplets pairing SRTM-derived DEMs with Sentinel-2 imagery and \emph{vision-grounded} natural-language captions that describe visible land cover. We position Geodiffussr as a strong baseline and step toward controllable 2.5D landscape generation for coarse-scale ideation and previz, complementary to physically based terrain and ecosystem simulators. We plan to release code and data to spur research on geometry-conditioned generative pipelines.
\end{abstract}

\ccsdesc[100]{Computing methodologies~Machine learning approaches}
\ccsdesc[100]{Computing methodologies~Neural networks}
\ccsdesc[300]{Computing methodologies~Machine learning}
\ccsdesc[500]{Computing methodologies~Texturing}
\ccsdesc[500]{Computing methodologies~Modeling and simulation}

%
%

\keywords{Terrain texturing, remote sensing, flow matching, multi-scale content aggregation}

\maketitle

\section{Introduction}
Realistic digital terrains are central to games, virtual production, simulation, and visualization. Practical pipelines must satisfy two competing demands: \emph{geometric fidelity} and \emph{visual richness} aligned with biomes or art direction. Manual authoring (sculpting, mask painting) is labor-intensive; procedural noise helps ideation but offers limited control when strict geometry adherence is required. In this work we explicitly target \emph{coarse-scale terrain ideation}: our experiments operate at 32$\times$32 base textures (\(\approx\)30\,m/px) and include 3D renders for communication.

We introduce \textit{Geodiffussr}, a text-guided, DEM-aware generative pipeline based on flow matching. The central design choice is \emph{multi-scale content aggregation} (MCA): VGG-derived DEM features are injected into UNet blocks at multiple resolutions, providing coarse-to-fine cues about global silhouettes and local ridge/valley structure. We also create a dataset pairing DEMs with satellite imagery and vision-grounded captions of visible land cover (appearance-centric rather than metadata-driven).

\paragraph{Positioning and scope.}
Geodiffussr is intended as a fast, controllable baseline for layout exploration and previz, complementary to physically-based terrain/biome stacks (erosion, sediment transport, snow/dune processes, ecosystem simulators). Our focus is adherence to a supplied DEM under text prompts; scaling to production resolutions is discussed in \S\ref{sec:discussion}.

In summary, we present the following contributions:
\begin{itemize}
    \item An open, biome-diverse remote sensing dataset containing triplets of DEMs, satellite images, and natural-language captions.
    \item A novel flow matching-based generative pipeline leveraging multi-scale content aggregation (MCA) for geometry adhering terrain texturing. Since this deals with a new task done in the text-to-terrain domain, we hope this work will stand as a baseline and support future works.
    \item Ablations showing a significant performance boost with our proposed method with MCA, as well as scaling with model size.
\end{itemize}

\begin{figure}[h]
  \centering
  \includegraphics[width=\linewidth]{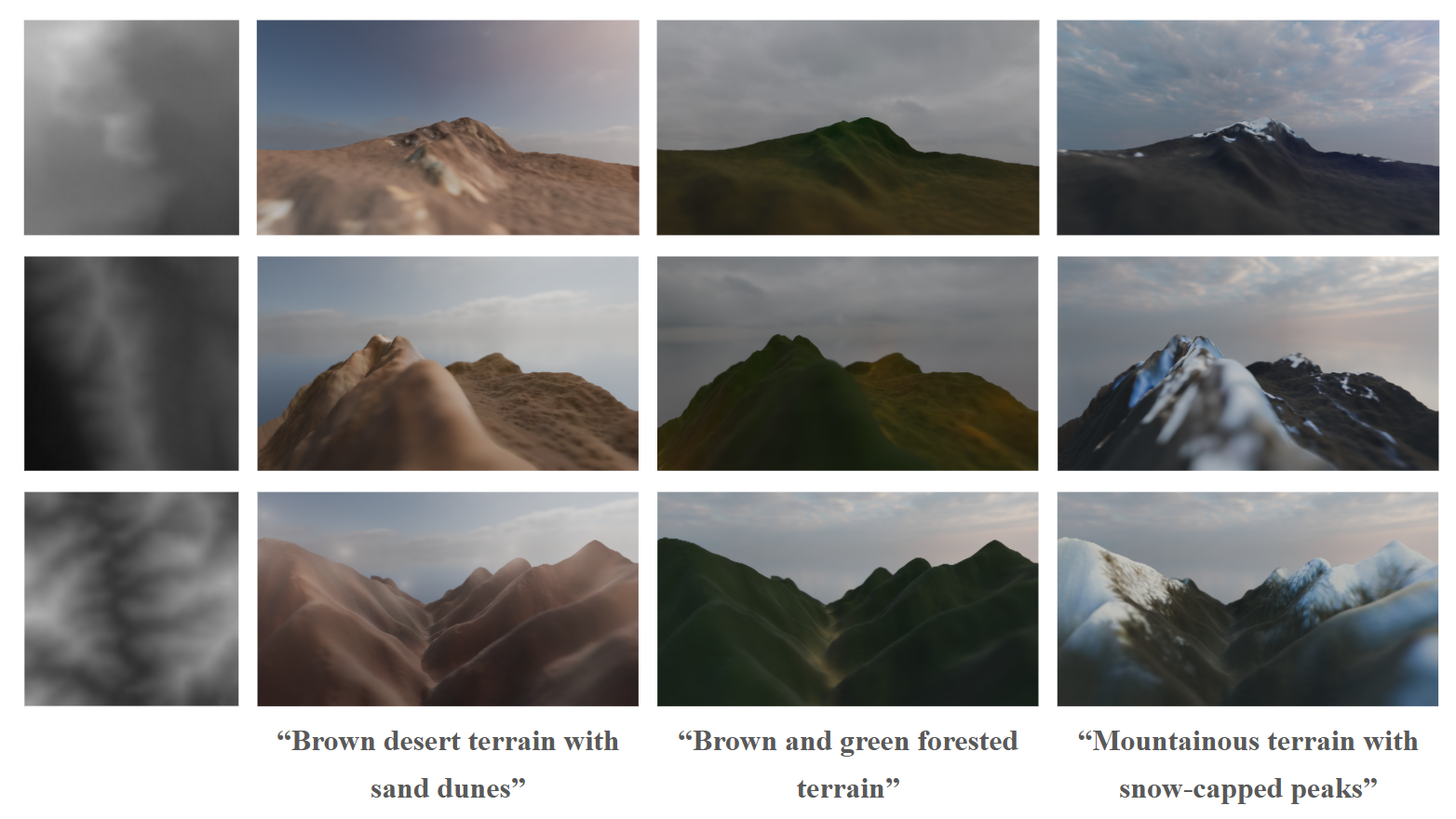}
  \caption{\textbf{Examples of rendered 2.5D terrains using our proposed approach.} We introduce Geodiffussr, a flow matching-based generative pipeline that can create terrain texture maps from intuitive text prompts, while realistically adhering to a specified Digital Elevation Map (DEM) by leveraging Multi-Scale Content aggregation (MCA). This provides a new baseline for text-conditioned, DEM-aware terrain synthesis and a stepping-stone toward fully controllable landscape generation.}
  \label{fig:conditioning}
\end{figure}

\begin{figure*}[t]
  \centering
  \includegraphics[width=\linewidth]{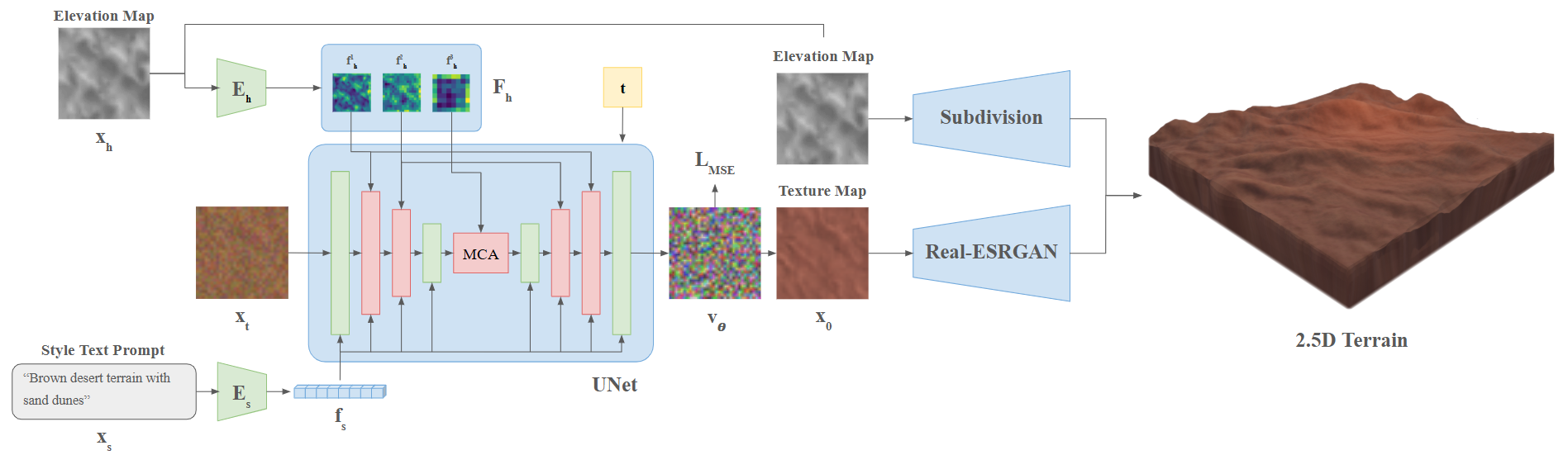}
  \caption{\textbf{Geodiffussr Pipeline.} We condition a flow matching model on both text embeddings and Digital Elevation Maps (DEMs). Specifically for DEMs, we take multi-scale features from a pretrained VGG-16 model and inject into the UNet blocks. The source DEM and generated texture map are increased in resolution via subdivision and Real-ESRGAN superresolution \cite{wang2021realesrgan_arxiv} respectively for rendering purposes. Combining these results in a 2.5D representation of a terrain as shown on the right.}
  \label{fig:modeldiagram}
\end{figure*}

\section{Related Work}
\noindent\textbf{Text-conditioned diffusion/flow.}
Diffusion/flow models achieve high-fidelity synthesis with conditioning (text, layout, edges); cross-attention provides semantics and auxiliary encoders inject structure.  

\noindent\textbf{Text-to-2.5D terrain.}
MESA \cite{BornePons_2025_MESA} trains a text-to-2.5D pipeline on global DEM–imagery pairs with metadata-driven captions. We instead use \emph{appearance-centric} captions and inject external DEM features via MCA to \emph{enforce} adherence to a provided DEM (we do not generate the DEM). 

\noindent\textbf{Physically-based terrain and ecosystems.}
Hydraulic/thermal erosion, sediment transport, snow/dune/glacier dynamics, and vegetation/ecosystem simulators deliver physical realism and layered materials \cite{Stava2008Hydraulic, Cordonnier2023GlacialErosion, Stomakhin2013SnowMPM}. Our learned, promptable texturing is complementary: it targets rapid, appearance-centric ideation conditioned on a supplied DEM and can be coupled with physical layers for production assets (\S\ref{sec:discussion}).

\section{Method}

\subsection{Model Architecture}
One of the most important aspects of the text-conditioned generation process is that it must adhere to the structure of the user-inputted height map. In order to enforce strict adherence to the height map, we incorporate Multi-scale Content Aggregation (MCA) \cite{Yang_2023_FontDiffuser} by injecting coarse-to-fine feature information into the UNet of the flow matching model. The motivation behind this conditioning mechanism is that image encoder's intermediate feature representations provide much richer information about the underlying height map than simply using the height map directly. In our implementation, we opted to use a pretrained VGG-16\cite{Simonyan2015} model and extract several feature maps at the 32x32, 16x16, and 8x8 resolutions and inject those into the Unet using a Squeeze-and-Excitation (SE)\cite{Hu2018} block. The SE block serves as a channel mixing mechanism to incorporate the information from the VGG feature maps before downsampling it back to the original size. 

For the text-conditioning we utilize text embeddings extracted from the final hidden state of the Flan-T5 series (Small, Base, and Large)\cite{Chung2022}, and perform pixel-wise cross attention in a similar fashion to popular diffusion models such as Stable Diffusion. Furthermore, we also incorporate pixel-wise self-attention blocks before performing the conditioning mechanisms. 

Figure~\ref{fig:modeldiagram} portrays the overall pipeline architecture detail, where a UNet model is conditioned on both text-embeddings and DEM features. Specifically, the DEM conditioning is done via MCA. After the Geodiffussr model, both DEM and texture are upscaled using simple subdivision and Real-ESRGAN superresolution \cite{wang2021realesrgan_arxiv} for 3D rendering purposes.



\subsection{Dataset}
We created a new dataset specialized for text-to-terrain purposes, containing 380K sets of Digital Elevation Maps (DEM), satellite images, and synthetic text labels.

For geospatially diverse data, we first constructed a catalogue of 200+ non-overlapping 1°$\times$1° Areas of Interest (AOIs) that jointly span every major terrestrial biome.
We performed biome stratification based on the WWF Terrestrial Ecoregions map and Koppen-Geiger climate classes \cite{Olson2001Ecoregions, Kottek2006KoppenGeiger}. For each of the 16 super-biomes  we targeted 10+ representative sites distributed across multiple continents. This was a method we implemented to obtain a smaller representative of the global geographical features without needing for the near to entire area coverage. Based on this catalogue, Digital Elevation Maps are sampled from USGS SRTMGL1 v003 \cite{farr2007srtm} and the satellite images are taken from Copernicus Sentinel-2 \cite{drusch2012sentinel2}.

Captions are synthesized using a pretrained language model \cite{Leenstra2021SelfSupervised}. Each satellite image is captioned using the Gemini 2.0 Flash Lite model\footnote{Google Gemini 2.0 Flash-Lite is a cost-efficient, low-latency variant of the Gemini 2.0 Flash family. Detailed specs and usage are available in the Vertex AI model garden: \url{https://cloud.google.com/vertex-ai/generative-ai/docs/models/gemini/2-0-flash-lite}}, considering its efficiency in multimodality. 

\section{Experiments}

\subsection{Ablation Study}
We isolate the effect of each design choice by varying one factor at a time while holding the rest fixed (AdamW, lr $5\!\times\!10^{-4}$, Flan-T5-Small for text, CFG scale $w{=}8$, etc.) 

Firstly, we observed the effect of MCA by comparing three settings: Full MCA ($32{\times}32$, $16{\times}16$, $8{\times}8$ VGG features with SE adapters), Single MCA (only $16{\times}16$ injection), and Non-MCA (direct DEM concatenation). Then, we compared model performance with its size, to see if there is potential for greater performance with increasing model size.


We evaluate Geodiffussr quantitatively and qualitatively, focusing on two axes: (i) texture quality, measured by FID and LPIPS, and (ii) elevation-texture alignment, measured by relative distance correlation ($\Delta$dCor) between the hue/saturation/value channels of the generated texture and the input DEM.

We report four complementary measures:
\begin{itemize}
  \item \textbf{FID} $\downarrow$ \cite{Heusel2017}: distributional distance between Inception features of generated vs.\ real satellite images (perceptual realism).
  \item \textbf{LPIPS} $\downarrow$ \cite{Zhang2018}: patch-level perceptual difference to reference imagery.
  \item \textbf{MSE} $\downarrow$: per-pixel reconstruction error (L$_2$).
  \item \boldmath $\Delta$\textbf{dCor} $\downarrow$: absolute gap to the dataset’s ground-truth dependence between DEM elevation and HSV$(X)$, i.e.,
  $\Delta\mathrm{dCor}=\big|\mathrm{dCor}(\mathrm{HSV}(X),\mathrm{DEM})-\mathrm{dCor}_{\text{gt}}\big|$,
  with $\mathrm{dCor}_{\text{gt}}{=}0.3816$ (captures geometry–appearance coupling, including nonlinear dependence).
\end{itemize}

\subsection{Main results}
Full MCA achieves FID 10.29, LPIPS 0.066, MSE 0.0166, and $\Delta\mathrm{dCor}$ 0.0016. Versus a non-MCA baseline, FID drops by 49.16\% and LPIPS by 32.33\%, while $\Delta\mathrm{dCor}$ improves from 0.0756 to 0.0016 (closing 97.9\% of the remaining gap to the dataset’s ground-truth dependence).

\subsection{Quantitative Effect of Multi-scale Content Aggregation (MCA)}

We compare three geometry injection settings: full MCA, single MCA, and non-MCA baseline (Table~\ref{tab:MCA}). When DEM features are fused at three scales (32×32, 16×16, 8×8) via MCA, the model attains its best scores across all metrics (FID 10.29, LPIPS 0.066, MSE 0.0166, and $\Delta$dCor 0.0016) indicating sharper, more geometry-consistent textures. Injecting at only the 16×16 scale yields intermediate performance (FID 14.50, LPIPS 0.085, MSE 0.0144, $\Delta$dCor 0.0196), while removing MCA entirely (direct DEM concatenation) degrades results (FID 20.24, LPIPS 0.0977, MSE 0.0184, $\Delta$dCor 0.0756) severely. 

What's especially notable is that full MCA injection closes 97.9 \% of the remaining dCor gap to the ground-truth, compared to a non-MCA baseline, demonstrating that multi-scale fusion enforces consistent shading and color transitions reflecting true topography present in the dataset. These findings confirm that multi-scale fusion substantially improves both perceptual fidelity and height–texture correlation.

 \begin{table}[ht]
\centering

\begin{tabular}{@{}p{0.20\columnwidth} p{0.15\columnwidth} p{0.15\columnwidth} p{0.15\columnwidth} p{0.15\columnwidth}@{}}
\toprule
Setting & FID $\downarrow$ & LPIPS $\downarrow$ & MSE $\downarrow$ & $\Delta$ dCor $\downarrow$ \\
\midrule
Full MCA & \textbf{10.29} & \textbf{0.066} & 0.0166 & \textbf{0.0016} \\
Single MCA & 14.50 & 0.085 & \textbf{0.0144} & 0.0196 \\
Non-MCA & 20.24 & 0.098 & 0.0184 & 0.0756 \\
\bottomrule
\end{tabular}
\vspace{1mm}
\caption{\textbf{Comparison between varying amounts of MCA injections.} Results show that injecting MCA into every dimension improves performance.\label{tab:MCA}}
\end{table}

\begin{figure}[h]
  \centering
  \includegraphics[width=1.0\linewidth]{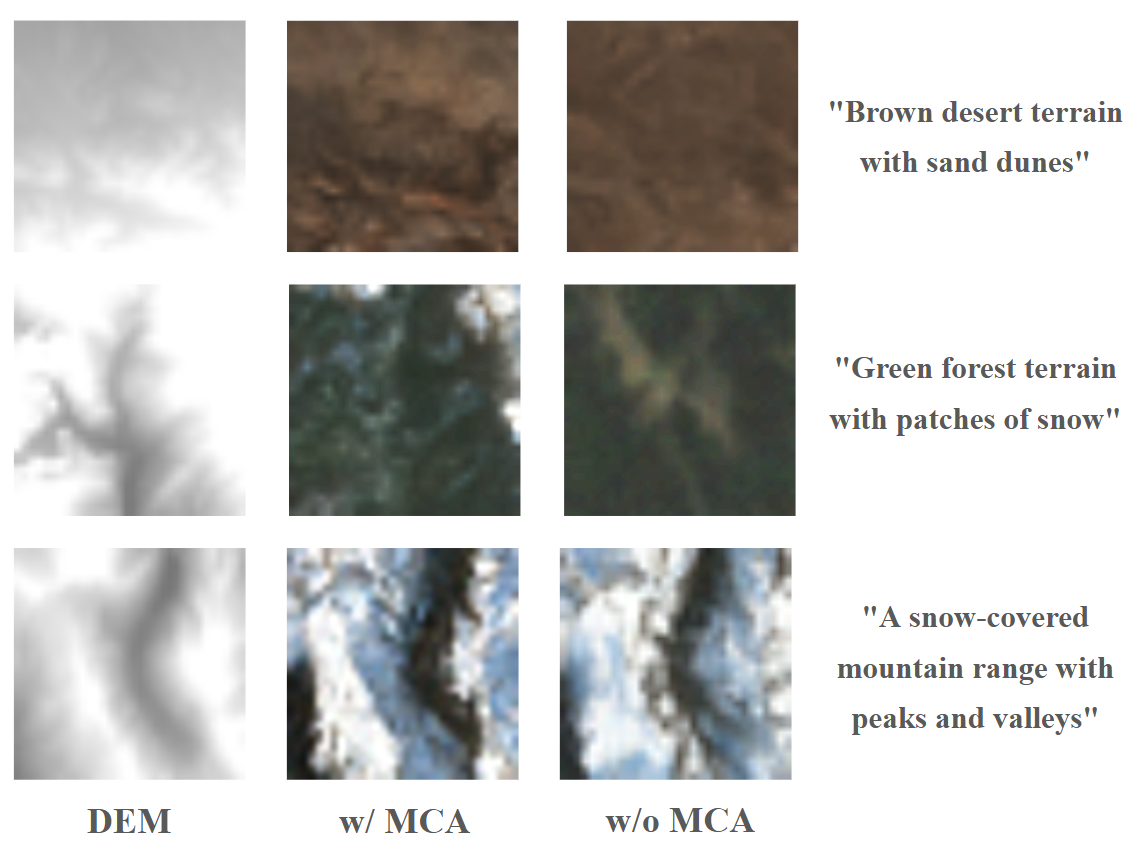}
  \caption{\textbf{Comparison of generated results between Full MCA (center) and non-MCA (right) versions of Geodiffussr.} The textures are generated with various prompts featuring different biomes and a source DEM (left).\label{fig:MCA}}
\end{figure}

\subsection{Qualitative Effect of Multi-Scale Content Aggregation (MCA)}

Figure~\ref{fig:MCA} contrasts outputs produced with and without Multi-scale Content Aggregation (MCA) when both models receive the same DEMs and prompts.

With MCA (center), the texture conforms to the underlying relief at every scale from global to local ones. For instance, with the snowy mountain range on the bottom row, snow settles cleanly on the ridge tops, darker rock appears at the valley, and subtle gray shadows accentuate minor spur lines. The viewer can infer the approximate height field from the texture alone, confirming that the network has internalized the DEM-to-texture relationship.

Without MCA (right), that correspondence collapses. Again with the same example, a single diagonal band of rock is hallucinated through the center while surrounding areas are indiscriminately coated in snow, ignoring the complex combination of peaks and valleys. 

This side-by-side shows that MCA’s coarse-to-fine fusion allows the network to internalize both global structural cues (e.g., mountain silhouettes) and fine-scale details (e.g., micro-ridges), whereas removing MCA entirely leads to severe texture collapse.

\subsection{UNet Model Size}
On the other hand, when comparing the performance of the generative models when varying the UNet model sizes, we observed a consistent increase in the performance of the models as the model capacity grew. These results reveal a promising trend that suggests future work may also benefit from scaling their models to even larger sizes than those we trained on, with little indication of a performance plateau within the 45M, 75M, 102M parameter models that we have tested. However, in line with established scaling laws, we expect that expanding model size will also require larger training sets to avoid diminishing returns.

\begin{table}[ht]
\centering
\begin{tabular}{@{}p{0.20\columnwidth} p{0.15\columnwidth} p{0.15\columnwidth} p{0.15\columnwidth} p{0.15\columnwidth}@{}}
\toprule
Model Size & FID $\downarrow$ & LPIPS $\downarrow$ & MSE $\downarrow$ & $\Delta$ dCor $\downarrow$ \\
\midrule
45M & 23.08	& 0.121 & 0.0235 & 0.0656 \\
75M & 14.50	& 0.085 & \textbf{0.0144} & 0.0196 \\
102M & \textbf{10.29} & \textbf{0.066} & 0.0166 & \textbf{0.0016} \\
\bottomrule
\end{tabular}
\vspace{1mm}
\caption{\textbf{UNet Model Size Comparison.} Results show that increasing model size steadily improves performance.\label{tab:unet}}
\end{table}

\begin{figure}[h]
  \centering
  \includegraphics[width=\linewidth]{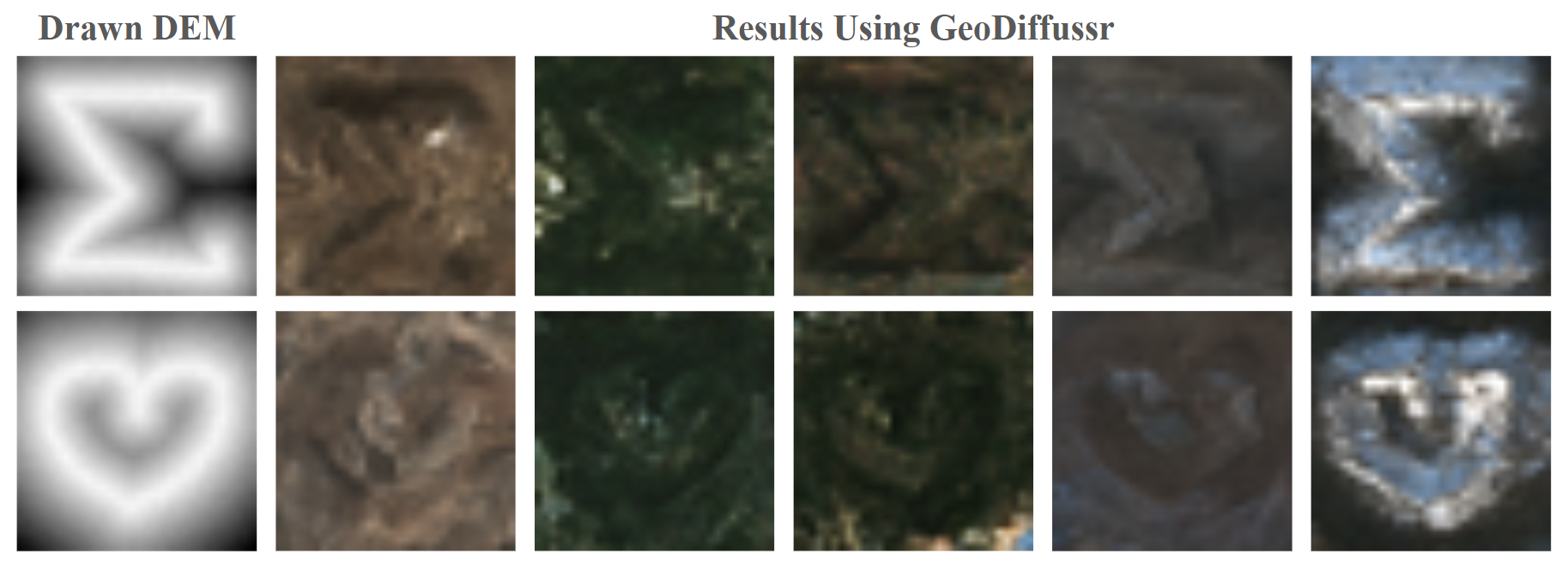}
  \caption{\textbf{Sketch DEMs.} Geodiffussr generalizes to user-drawn synthetic DEMs, producing coherent, prompt-consistent textures. This demonstrates the flexibility of our model to unseen complex geometry, and its potential to be applied with user-guided DEMs.}
  \label{fig:symbols}
\end{figure}

\section{Discussion and Limitations}\label{sec:discussion}
\textbf{Why MCA works.} Injecting DEM features at multiple scales exposes the UNet to both global silhouettes and fine relief, which we find is essential for consistent shading/biome placement relative to elevation.


\textbf{Scaling to production.}
Since Geodiffussr is still working with a coarse resolution ($32{\times}32$), we suggest some approaches to higher resolutions for practical use:
(i) a \emph{global-context token} from pooled full-scene DEM features injected into MCA at each scale to keep long-range structure coherent,
(ii) a \emph{coarse-to-fine cascade}  (32$\rightarrow$128$\rightarrow$256\,px) conditioned on the upsampled prior stage with elevation-aware regularizers (e.g., gradient alignment) to preserve snowlines and drainage, and
(iii) a \emph{DEM-aware super-resolution head} tuned for geospatial edges.

\textbf{Applications and integration.} We envision an end-to-end 2.5D pipeline driven by text and sketches: a sketch-to-DEM module converts hand-drawn contours into elevation maps \cite{Wang2020, Hu2024}, and Geodiffussr applies multi-scale, promptable texturing—yielding terrains whose elevation and appearance jointly follow the user’s sketch and prompt. A prototype of this idea is illustrated in Figure~\ref{fig:symbols}

\section{Conclusion}
Geodiffussr couples text guidance with explicit, multi-scale DEM conditioning for terrain texturing. Our experiments validate MCA’s impact on perceptual quality and elevation alignment, establishing a compact, reproducible baseline for controllable 2.5D terrain synthesis. We hope to openly contribute our baseline and dataset to spur future research into realistic, user-guided terrain texturing.



\end{document}